\documentclass[aps,prb,twocolumn,superscriptaddress]{revtex4}
\usepackage{graphicx} %graphics handler
\usepackage{dcolumn}% Align table columns on decimal point
\usepackage{bm}% bold math

\begin{document}

\title{Rapid Scanning Terahertz Time-Domain Magnetospectroscopy \\with a Table-Top Repetitive Pulsed Magnet}
 
\author{G.~Timothy Noe}
\author{Qi Zhang}
\author{Joseph Lee}
\affiliation{Department of Electrical and Computer Engineering, Rice University, Houston, TX 77005, USA}

\author{Eiji Kato}
\affiliation{Advantest America, Inc., Princeton, New Jersey, USA}

\author{Gary L.~Woods}
\affiliation{Department of Electrical and Computer Engineering, Rice University, Houston, TX 77005, USA}

\author{Hiroyuki Nojiri}
\affiliation{Institute for Materials Research, Tohoku University, Sendai, Japan}

\author{Junichiro Kono}
\email[]{kono@rice.edu}
\thanks{corresponding author.}
\affiliation{Department of Electrical and Computer Engineering, Rice University, Houston, TX 77005, USA}
\affiliation{Department of Physics and Astronomy, Rice University, Houston, TX 77005, USA}
\affiliation{Department of Materials Science and NanoEngineering, Rice University, Houston, TX 77005, USA}

\date{\today}

\begin{abstract}
We have performed terahertz time-domain magnetospectroscopy by combining a rapid scanning terahertz time-domain spectrometer based on the electronically coupled optical sampling method with a table-top mini-coil pulsed magnet capable of producing magnetic fields up to 30~T.  We demonstrate the capability of this system by measuring coherent cyclotron resonance oscillations in a high-mobility two-dimensional electron gas in GaAs and interference-induced terahertz transmittance modifications in a magnetoplasma in lightly doped $n$-InSb.
\end{abstract}

%\pacs{78.67.Ch,71.35.Ji,78.55.-m}
%\keywords{}

\maketitle


\section{Introduction}

There are a number of low-energy excitations, both of spin and orbital origins, in solids placed in strong magnetic fields.\cite{PetrouMcCombe91LLS,BoebingeretAl01Book,KonoMiura06HMF,Miura07Book}  These excitations, typically occurring in the terahertz (THz) frequency range, are ideally suited for fundamental studies of quantum coherent dynamics of non-equilibrium many-body states as well as for emerging quantum technologies such as spintronics.\cite{Kono11NP}  However, there has been only limited success in combining THz time-domain spectroscopy (THz-TDS) techniques with high magnetic fields.\cite{SomeNurmikko94APL,HuggardetAl97APL,KonoetAl99APL,Crooker02RSI,WangetAl07OL,WangetAl10NP,WangetAl10OE,MolteretAl10OE,ArikawaetAl11PRB,MolteretAl12OE}  In particular, combining THz-TDS with a pulsed magnet remains to be a significant technical challenge,\cite{Crooker02RSI,MolteretAl10OE,MolteretAl12OE} while magnetic fields stronger than 45~T can be generated only in pulsed form.\cite{MiuraHerlach85Book}

The traditional method for measuring a THz time-domain waveform generated using ultrashort laser pulses includes using a pump-probe scheme where one laser beam passes a beam splitter to make two synchronized pulses from one source.  One of the pulses, the pump, is used to generate a THz pulse with duration on the order of picoseconds and the other pulse is used to probe the THz electric field at a given time delay after the THz radiation interacts with the sample.  This method traditionally incorporates a mechanical translation stage and retroreflector mirrors to control the relative timing between the probe pulse and the THz radiation.  Efficient alternatives to step-scan delay stage techniques for ultrafast time-resolved spectroscopy include asynchronous optical sampling (ASOPS)\cite{ElzingaetAl87AO} and electronically controlled optical sampling (ECOPS).\cite{KimetAl10OL}  These methods use two ultrafast lasers with a repetition rate offset, $\Delta f$, between a master laser (with repetition rate, $f_\mathrm{Ref}$) and a slave laser (with repetition rate, $f_\mathrm{Ref} \pm \Delta f$) so that for each pair of pulses exiting the lasers, the relative timing of the pulses changes.  In the ASOPS technique, the two lasers have a fixed repetition rate offset such that one of the laser's pulses `catch up' with the other's after a measurement time of $1/\Delta f$.  The full range of time delay for an ASOPS setup is $1/f_\mathrm{Ref}$ with time delay resolution of $\Delta f/f_\mathrm{Ref}^{2}$.  For example, with a master repetition rate of 100~MHz and a repetition rate offset of 1~kHz, 10~ns of pump-probe time delay can be scanned in 1~ms with time delay resolution of 100~fs.  The ECOPS technique modulates the repetition rate offset such that the time delay of the probe pulses relative to the pump pulses is modulated in a small time delay region around the time zero point.  For time-resolved measurements such as THz-TDS, where the only interesting features occur in a small time range compared to the full range, $1/f_\mathrm{Ref}$, the ECOPS method makes more efficient use of measurement time.

Recently, we have developed a 30~T mini-coil pulsed magnet specifically for nonlinear and/or ultrafast spectroscopy.\cite{NoeetAl13RSI}  Our magneto-optical system combines the following experimental conditions: sample temperatures down to $\sim$10~K, magnetic fields up to 30~T, and direct optical access via windows on either side of the dual cryostat system.  Previous work combining THz-TDS with high magnetic fields (above the strength achieved with commercially available magnets with direct optical access) include inserting fiber coupled emitters and detectors into the bore of a direct current (DC) superconducting magnet\cite{Crooker02RSI} or into the cryostat of a large capacitor bank driven pulsed magnet;\cite{MolteretAl10OE} they used a mechanical translation stage and a rotating delay line, respectively. Other related work includes using a smaller, lower magnetic field capability (up to 6~T) pulsed magnet, where the THz time-domain waveform is measured utilizing a step-scan method with a mechanical translation stage; the magnetic field dependence of the THz electric field is recorded for each time delay, and after a series of magnetic field pulses, each with different probe time delay, the waveforms can be recorded.\cite{MolteretAl12OE}  In the present work, we couple an ECOPS-based THz-TDS spectrometer with our pulsed magnet via free-space optics to efficiently perform THz-TDS measurements at the peak of the magnetic field.

\section{METHODS}
\label{methods}

%%%%% Fig. 1 %%%%%
\begin{figure}
\begin{center}
\includegraphics[scale=0.42]{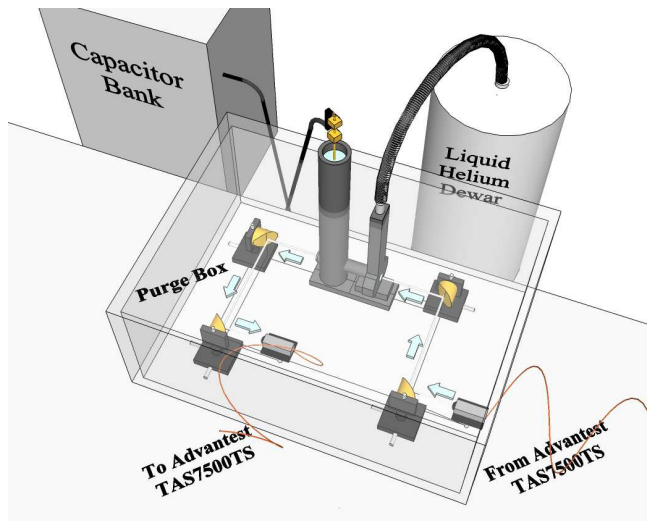}
\caption{Optical setup.  The magnet is incorporated into a transmission THz-TDS optical path under a dry nitrogen purged environment.  The THz emitter and detector are coupled to the dual-laser system via optical fibers and electrical wires.  Two wire-grid polarizers are located before and after the magnet with parallel transmission axis.}
\label{ill:OptSetupFigure}
\end{center}
\end{figure}
%%%%%%%%%%

The commercial ECOPS-based THz-TDS spectrometer (Advantest Corporation, TAS7500TS) that we used for our measurements incorporates a photoconductive antenna THz emitter and detector coupled with dual ultrashort pulse fiber lasers, which have a center wavelength of 1.55~$\mu$m, via optical fibers and electrical wires that are housed in a mountable modules with silicon lenses.  For our measurements, we incorporated the emitter and detector modules into a transmission geometry beam path, where one off-axis parabolic mirror collimates the THz radiation from the emitter, the next focuses it through our magnet system, the next re-collimates it, and, finally, the last focuses it onto the detector (see Fig.~\ref{ill:OptSetupFigure}).

%%%%% Fig. 2 %%%%%
\begin{figure*}
\begin{center}
\includegraphics[scale=0.7]{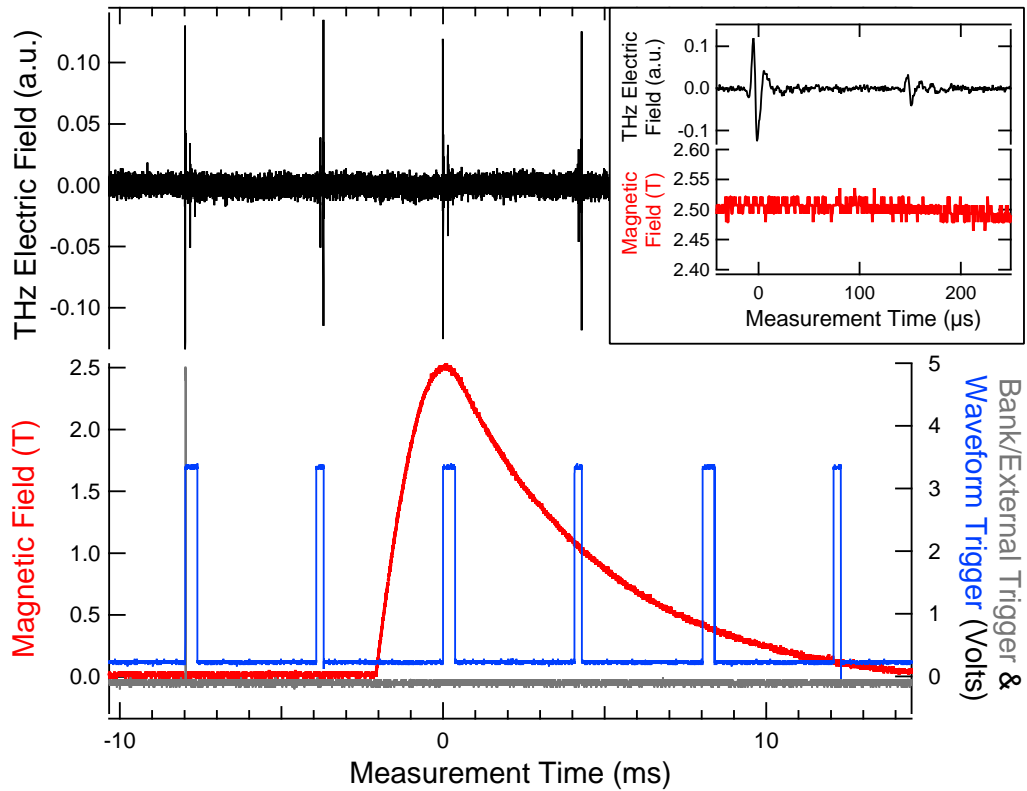}
\caption{Synchronization timing scheme.  The THz spectrometer sends a trigger signal to the magnet control unit. The generation of the magnetic field is delayed 6~ms such that the next forward sweeping THz waveform readout occurs at the peak of the magnetic field.  The inset shows the data quality for the THz waveform signal for a single sweep at the peak of the magnetic field where the magnetic field variation is less than 1\%.}
\label{ill:SyncFigure}
\end{center}
\end{figure*}
%%%%%%%%%%

Two wire-grid polarizers with parallel transmission axis were used to ensure that we were measuring the response of linearly polarized THz radiation through our samples. Relative to our previous measurements\cite{NoeetAl13RSI} using the mini-coil pulsed magnet, we increased the numerical aperture from the helium-flow cryostat side to 0.06 by increasing the size of the tapped hole on the helium-flow cryostat's cold finger, screwing a copper pipe cold finger extension to the cryostat's cold finger, and securing a shorter sapphire pipe to this extension.  This modification allowed us to increase the amount of THz radiation that transmits through the coupled cryostats.

We synchronized the generation of the magnetic field pulse relative to the readout of THz waveform by delaying the start of the magnetic field pulse by 6~ms with respect to an external trigger signal that occurs during a forward-sweeping THz waveform readout.  The master laser of the THz spectrometer had a nominal repetition rate of 50~MHz.  The repetition rate of the slave laser was modulated with respect to the master laser's repetition rate such that two consecution forward sweeping readouts of the THz waveform were separated by 8~ms with a backward sweeping readout in between (see Fig.~\ref{ill:SyncFigure}).  Because the time between the start and peak of the magnetic field was 2~ms, a forward-sweeping THz waveform readout occurred during the time of the peak magnetic field.  We triggered the oscilloscope (Tektronix, DPO 7254) on the second waveform trigger signal after the bank/external trigger.  The inset of Fig.~\ref{ill:SyncFigure} shows an expanded view of the THz waveform signal for a single readout sweep at the peak of the magnetic field where the magnetic field strength is relatively constant, less than 1\%, between the main pulse and the reflected pulse of the back of the sample.  Because the emitter and detector were housed in packaged modules and the full waveform of THz signal from the ECOPS-based spectrometer could be monitored in real time, maximizing the THz radiation through THz beam path and through the coupled cryostats was simple and straightforward.

\section{RESULTS}

\subsection{Coherent Cyclotron Resonance Oscillations in a Two-Dimensional Electron Gas}

%%%%% Fig. 3 %%%%%
\begin{figure*}
\begin{center}
\includegraphics[scale=0.65]{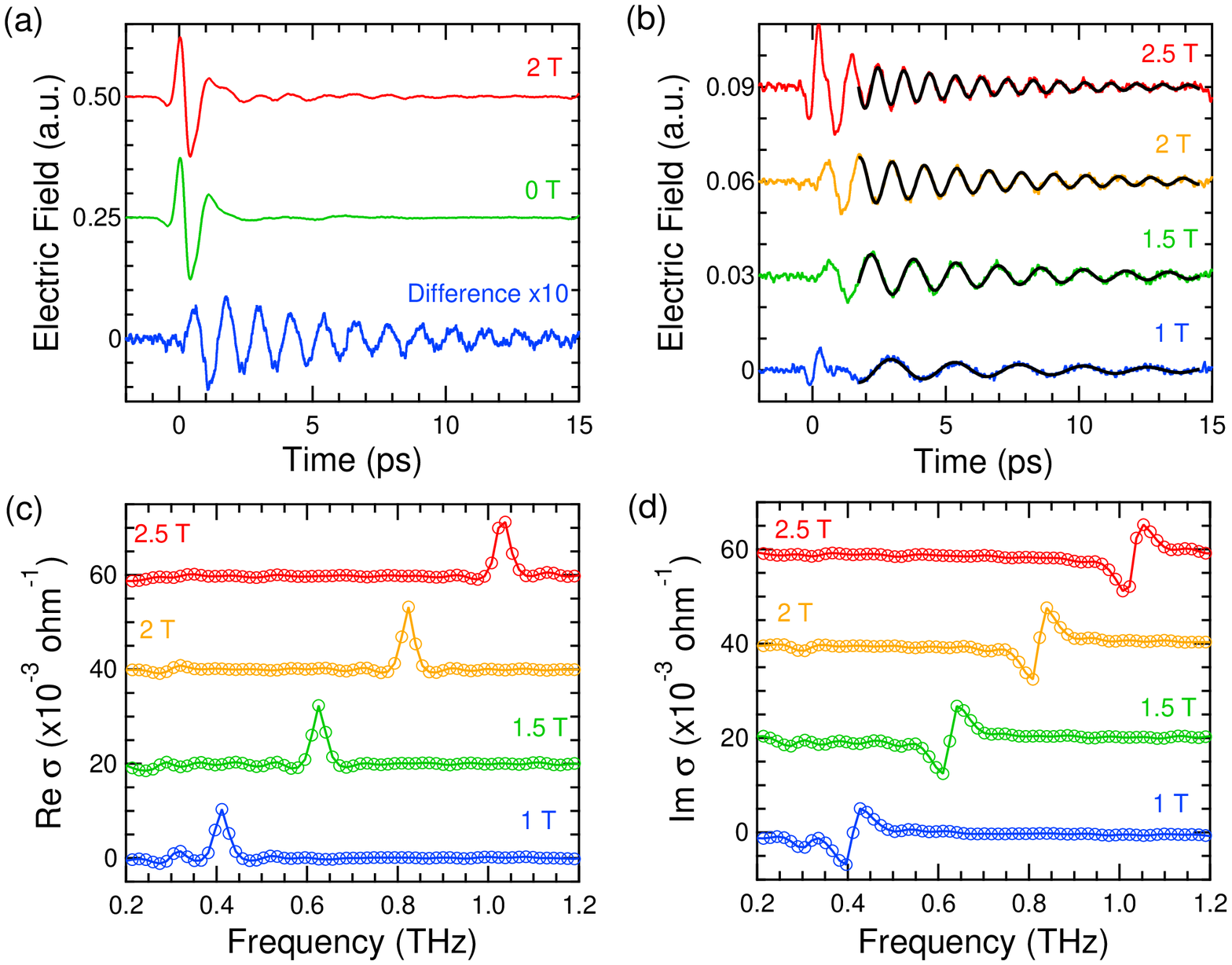}
\caption{Coherent cyclotron resonance (CCR) observed for a two-dimensional electron gas in GaAs.  \textbf{(a)}~Example of time-domain THz waveforms taken after 64 averages at 2~T and 256 averages at 0~T.  The difference reveals CCR oscillations.  \textbf{(b)}~CCR oscillation data with an exponentially damped sine wave fit at various magnetic fields.  \textbf{(c)}~The real part and \textbf{(d)}~imaginary part of the conductivity at various magnetic fields. The sample temperature was $\sim$10.5~K.}
\label{ill:2DEG_Figure}
\end{center}
\end{figure*}
%%%%%%%%%%

We measured coherent cyclotron resonance (CCR) oscillations in a two-dimensional electron gas (2DEG) sample with a 4.2-K mobility of 2.2 $\times$ 10$^6$~cm$^2$/Vs and density of $\sim$3 $\times$ 10$^{11}$~cm$^{-2}$, similar to those studied in previous measurements utilizing a commercial 10-T DC superconducting magnet.\cite{WangetAl07OL,WangetAl10OE,ArikawaetAl11PRB} Figure~\ref{ill:2DEG_Figure}\textbf{a} shows the THz time-domain waveform at 2~T after 64 averages and 0~T after 256 averages.  Subtracting the 0~T result from the 2~T result reveals CCR oscillations clearly visible in the time domain.  The CCR oscillations take the functional form of an exponentially decaying sine wave,
\begin{equation}
CR(t)=Ae^{-t/\tau_\mathrm{CR}}\sin(\omega_\mathrm{c}t+\phi_{0}),
\label{CR(t)}
\end{equation}
where $A$ is the amplitude, $\tau_\mathrm{CR}$ is the CR decay time, $\omega_\mathrm{c}$ = $eB/m^*$ is the cyclotron frequency, $e$ is the electron charge, $B$ is the magnetic field, $m^{*}$ is the effective mass of the electrons, and $\phi_{0}$ is the phase offset.  Figure~\ref{ill:2DEG_Figure}\textbf{b} shows the CCR oscillations for magnetic field strengths of 1~T, 1.5~T, 2~T, and 2.5~T along with the fitting results from Eq.~(\ref{CR(t)}). The results for the frequency, $\nu_\mathrm{c} = \omega_\mathrm{c}/2\pi$, are 0.42~THz, 0.63~THz, 0.82~THz, and 1.03~THz, respectively, which, after fitting $\nu_\mathrm{c}$ vs.~$B$ with a straight line, yield an effective mass, $m^{*}$ = 0.068$m_{0}$, where $m_{0}$ is the free electron mass.  The results for the CR decay time, $\tau_\mathrm{CR}$, are 7.5~ps, 6.4~ps, 6.0~ps, and 5.3~ps, respectively.

By taking the ratio of the Fourier transforms of the time-domain data for the transmitted THz electric field at finite $B$ and $B$ = 0, we can calculate the complex conductivity, $\tilde{\sigma}(\nu,B)$, of the 2DEG.  The equation for the ratio of the Fourier transforms in the thin-film approximation is\cite{NussOrenstein98THz}
\begin{equation}
\frac{\tilde{E}(\nu,B)}{\tilde{E}(\nu,B=0)}=\frac{\frac{1+n}{Z_{0}}}{\frac{1+n}{Z_{0}}+\tilde{\sigma}(\nu,B)},
\end{equation}
where $\tilde{E}(\nu,B)$ is the complex THz electric field in the frequency domain transmitted through the sample at finite $B$, $\tilde{E}(\nu,B=0)$ is the complex electric field in the frequency domain transmitted through the sample at $B=0$, $n$ is the index of refraction for the GaAs substrate, and $Z_{0} = 377$~$\Omega$ is the impedance of free space. Note that $\tilde{\sigma}(\nu,0)$ $\approx$ 0 in the THz frequency range for high-mobility 2DEGs.  The calculated results for the real and imaginary parts of the complex conductivity are shown in Figure~\ref{ill:2DEG_Figure}\textbf{c} and \ref{ill:2DEG_Figure}\textbf{d}, respectively.

\subsection{Interference-induced effects in an $n$-InSb semiconductor magnetoplasma}

%%%%% FIG. 4 %%%%%
\begin{figure*}
  \centering
  \includegraphics[scale=0.6]{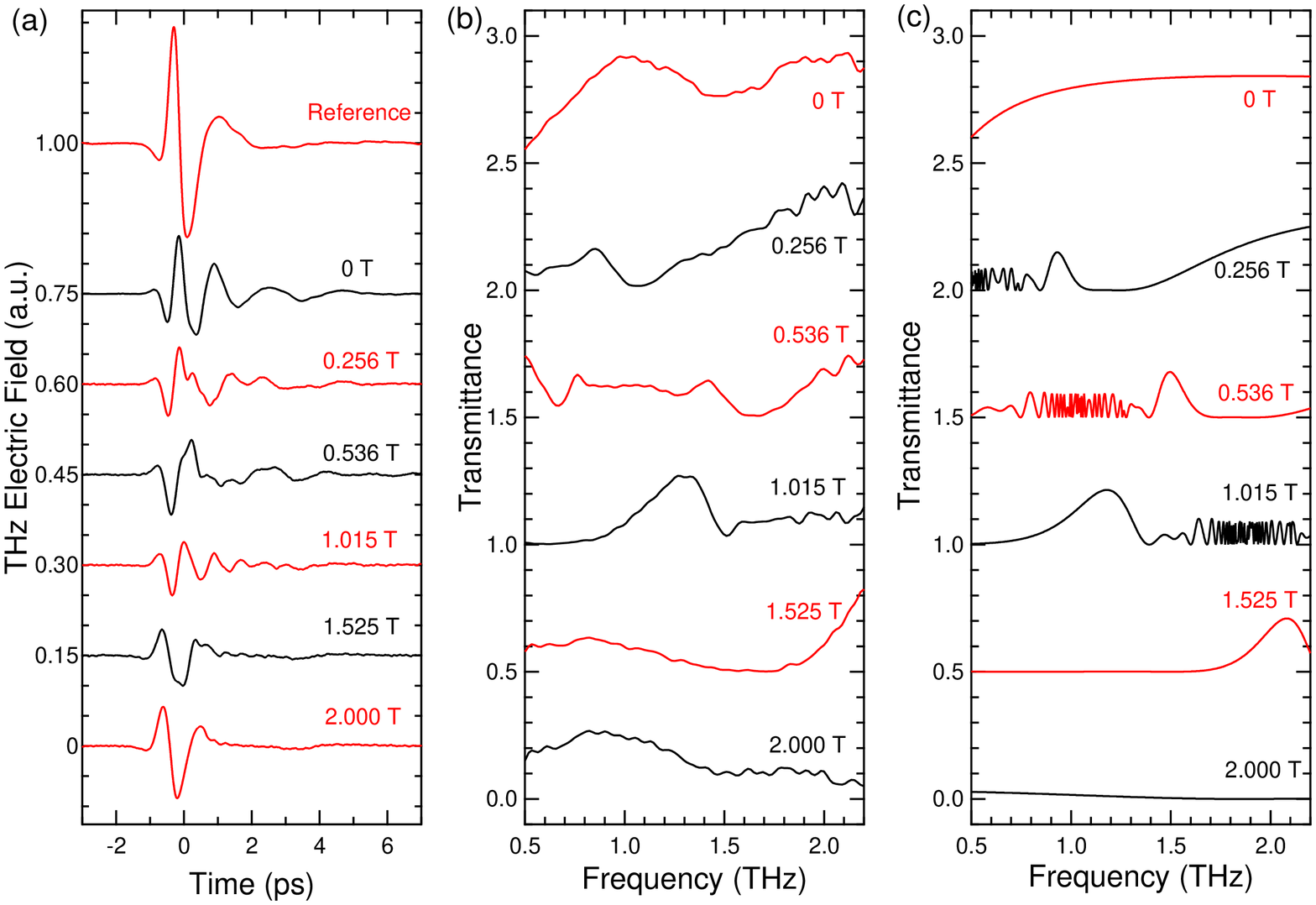}
\caption{\textbf{(a)}~Transmitted time-domain waveform after 64 averages through $n$-InSb at various magnetic fields at 41~K and the reference aperture waveform after 256 averages.  \textbf{(b)}~Measured and \textbf{(c)}~simulated transmittance of THz radiation though $n$-InSb as a function of frequency at various magnetic fields at 41~K.}
\label{ill:InSb_Figure}
\end{figure*}
%%%%%%%%%%%%%%%

We measured the transmittance of bulk $n$-InSb with a density of 3.5 $\times$ 10$^{14}$~cm$^{-3}$, similar to previous measurements with a commercial 10-T DC superconducting magnet.\cite{WangetAl10NP}  This sample has a plasma frequency of $\sim$0.3~THz at 0~T and a cyclotron frequency of $\sim$2~THz at 1~T.  In the presence of a magnetic field, the plasma edge splits into two, one for electron-CR-active (CRA) and the other for electron-CR-inactive (CRI) circular polarizations.  A linearly polarized THz beam incident onto the sample is a 50\%-50\% mixture of CRA and CRI polarized waves, which, upon entering into the crystal, propagate at different, $B$-dependent phase velocities.  Thus, after propagating through the entire sample, they have different phases, resulting in complicated $B$-dependent interference patters in transmittance spectra, which can be simulated through a classical magnetoplasma model.\cite{WangetAl10NP}

Figure~\ref{ill:InSb_Figure}\textbf{a} shows the THz time-domain waveform transmitted through the sample at various magnetic fields and 41~K to be compared to the reference waveform taken transmitting through the cryostats without a sample.  The time delay of the reference has been offset relative to the other traces as it would arrive at an earlier time because of the index of refraction of the sample.  For each increasing magnetic field strength, the time-domain waveform dramatically changes until the 2~T strength, where it resembles the shape of the reference waveform.  The transmittance of the InSb sample can be obtained by taking the Fourier transform of the time-domain waveform at various magnetic fields in the time period around the main pulse and dividing by the Fourier transform of the reference.  Figure~\ref{ill:InSb_Figure}\textbf{b} shows the experimental results for the transmittance spectra to be compared with the simulated transmittance spectra, Figure~\ref{ill:InSb_Figure}\textbf{c}, obtained using the method outlined in Reference \onlinecite{WangetAl10NP}.  The transmittance of the magnetoplasma sensitively changes with increasing magnetic field and the main features of the transmittance, peaks, plateaus, and valleys, are reproduced by the simulations.

\section{CONCLUSION}

We have demonstrated the ability to perform THz time-domain magneto-spectroscopy at the peak of the generated magnetic field using our mini-coil pulsed magnet capable of producing magnetic field strengths of 30~T in a straightforward manner utilizing the ECOPS approach to ultrafast spectroscopy.  As an example, we observed coherent cyclotron resonance oscillations in a high-mobility two-dimensional electron gas in GaAs and interference-induced terahertz transmittance modifications in a magnetoplasma in lightly doped $n$-InSb.  Potentially, one could maximize the efficiency of these types of measurements to capture the full magnetic field dependence from 0 to 30~T by utilizing a very high-speed ASOPS based THz-TDS spectrometer,\cite{BartelsetAl07RSI} where on the order of 100~THz waveforms could be recorded during the magnetic field pulse instead of only at the peak of the field with little magnetic field variation for each waveform readout.  With such a setup, one could extremely efficiently obtain the THz response of a sample as a function of magnetic field in one magnetic field shot.  Depending on the signal-to-noise ratio of each THz waveform readout, multiple magnet shots may be required to average the signal to provide high-quality two-dimensional image maps of, for example, transmittance as a function of THz frequency vs.~magnetic field.

%% classes.  Rather than use \section, an appropriate macro is
%% provided that will always work.
%%%%%%%%%%%%%%%%%%%%%%%%%%%%%%%%%%%%%%%%%%%%%%%%%%%%%%%%%%%%%%%%%%%%%
\section*{Acknowledgements}

This work was supported by the National Science Foundation through Grants No.~DMR-1006663 and No.~OISE-0968405.  We thank %Advantest America, Inc., for their generosity and efforts in loaning us the THz-TDS system used to perform these measurements, 
John Reno at Sandia National Laboratories and CINT for providing us with the high-mobility 2DEG sample, Scott Crooker at Los Alamos National Laboratory for the provision of the InSb sample, Alexey Belyanin for the computer program used to simulate the transmittance spectra for InSb, and Trevor Smith for his assistance in performing preliminary measurements.

%\pagebreak

%\bibliography{jun}

\end{document}